\documentclass[a4paper]{article}

\usepackage[utf8]{inputenc}
\usepackage[T1]{fontenc}

\usepackage{XCharter}
\usepackage[charter,scaled=1.1]{newtxmath}
\usepackage{microtype}

\usepackage[british]{babel}
\usepackage[sorting=none,bibstyle=numeric,citestyle=numeric-comp]{biblatex}
\usepackage{csquotes} 
\usepackage{hyperref}
\usepackage{graphicx}
\usepackage{booktabs}
\usepackage{mathtools} 

\newcommand{\bigO}{\mathcal{O}}
\newcommand{\diff}{\mathrm{d}}
\newcommand{\half}{\frac{1}{2}}

\AtBeginBibliography{\raggedright}

\title{Co-simulation errors due to step size changes}
\author{Lars T. Kyllingstad\footnote{%
    SINTEF Ocean, Trondheim, Norway.
    \href{mailto:lars.kyllingstad@sintef.no}{\textit{lars.kyllingstad@sintef.no}}.
    \href{https://orcid.org/0000-0002-4334-4490}{ORCID 0000-0002-4334-4490}.
}}
\date{}

\begin{document}
\maketitle

\begin{abstract}
    When two simulation units in a continuous-time co-simulation are connected
    via some variable $q$, and both simulation units have an internal state
    which represents the time integral of $q$, there will generally be a
    discrepancy between those states due to extrapolation errors.
    Normally, such extrapolation errors diminish if the macro time step size is
    reduced.
    Here we show that, under certain circumstances, step size changes can cause
    such discrepancies to \emph{increase} even when the change is towards
    smaller steps.
\end{abstract}

\section{Introduction}

Co-simulation is an approach to simulation of coupled dynamical systems
where a system of interest is divided into subsystems, each of which is
simulated independently of the others, only exchanging information at discrete
communication points.
Between communication points, each such \emph{simulation unit}
is responsible for evolving its own internal state forward in time,
for example by integrating a set of ordinary differential equations.

This decoupling of subystems makes co-simulation especially well suited for
multiphysical simulation, distributed computing, and, in an industrial
setting, protection of sensitive information about individual subsystems
\cite{gomes2018cosimulation}.
However, it is also a source of errors.
Since simulation units only sample the true values of their inputs at
communication points, they have to approximate them during the intervening time
intervals.
With explicit co-simulation methods, simulation units must extrapolate from past
input values, and with implicit methods, they can interpolate input variables
between one communication point and the next.
Such approximations are by nature inaccurate and can lead to discontinuities at
communication points.
Thus, co-simulation introduces \emph{coupling errors} on top of errors already
present in the numerical computation of subsystem behaviours.

As long as the simulation units are convergent and have continuous behaviour,
coupling errors scale as some power of the temporal distance
between communication points
\cite{kubler2000two,arnold2001preconditioned,arnold2013error,glumac2023defect},
called the \emph{macro time step size}.\footnote{
    The prefix ``macro'' is used to distinguish the time steps between
    communication points from the internal (micro) time steps used within
    simulation units.
    In this paper we will not refer to micro time steps at all, and will
    therefore often drop the prefix.
}
This means that one can reduce errors by reducing the step size, at the cost of
increased computational time.
This is used to good effect in variable-step-size co-simulation algorithms,
which dynamically adjust the step size during the course of a
simulation to maximise performance while keeping errors within certain bounds
\cite{busch2011explicit,schierz2012cosimulation,sadjina2017energy,meyer2021cosimulation,eguillon2022f3ornits,glumac2023defect,kyllingstad2025error}.

However, as we will show in this paper, in some cases the very act of changing
the step size can induce errors, even when the change is towards shorter steps.
Specifically, we will show that when a coupling variable is integrated on both
sides of a connection, so that the integral on one side is based on an
approximation, the resulting coupling errors are
proportional to the \emph{difference} between subsequent step sizes.

One can argue that having two instances of the same state in one simulation is
redundant and should be avoided.
For the sake of numerical accuracy, it would be better to maintain the
integrated state in just one simulation unit, and expose it as an output so that
other simulation units can access its ``true'' value.
In co-simulation, however, it often makes perfect sense to make simulation
units as self-contained as possible, avoiding structural dependencies on other
simulation units, to ensure modularity, reusability, and scalability.
Presenting a minimal interface against other simulation units is part of this
\cite{kyllingstad2025error}.

The described situation typically occurs when a \emph{flow} variable -- for
example velocity, angular velocity, electrical current, or volumetric flow
rate -- is exchanged between simulation units and integrated to produce a
\emph{displacement} -- position, angle, electric charge, or displaced volume,
respectively.
In this paper, we will present two such examples, in sections
\ref{sec:harmonic-oscillator} and \ref{sec:connected-reservoirs}, as well as a
general derivation of the phenomenon in \autoref{sec:general-case}.
We conclude the paper with some observations on how the described errors can be
mitigated in \autoref{sec:conclusion}.

\section{Example 1: Damped harmonic oscillator}
\label{sec:harmonic-oscillator}

\begin{figure}
    \centering
    \includegraphics{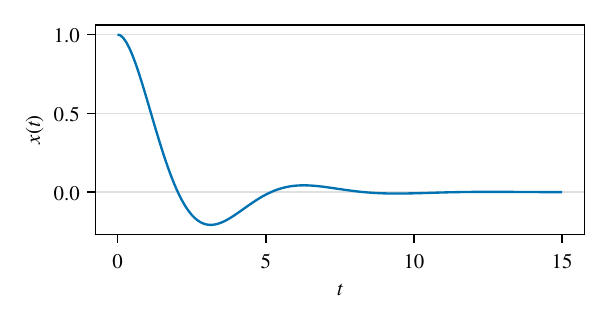}
    \caption{Solution of the damped harmonic oscillator with $m = c = k = 1$ and
        initial conditions $x(0) = 1$ and $\dot x(0) = 0$.}
    \label{fig:harmonic-oscillator-accurate}
\end{figure}

As our first example, we consider the case of a damped harmonic oscillator:
\begin{equation}
    m \ddot x(t) + c \dot x(t) + k x(t) = 0.
\end{equation}
For the sake of clarity, we shall work in dimensionless units and take the mass,
damping coefficient, and spring constant to be unity: $m = c = k = 1$.
\autoref{fig:harmonic-oscillator-accurate} shows the solution for initial
conditions $x(0) = 1$ and $\dot x(0) = 0$.

\begin{figure}
    \centering
    \includegraphics{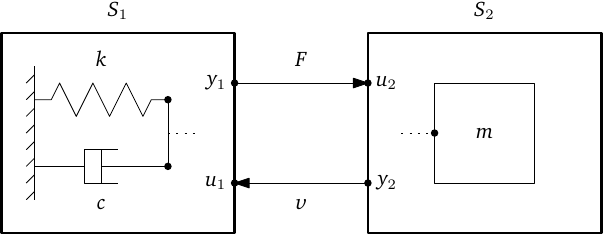}
    \caption{Co-simulation setup of the damped harmonic oscillator.}
    \label{fig:mass-spring-damper-cosim}
\end{figure}

For the purpose of co-simulation, we envision it in terms of a
mass--spring--damper system and split it into two simulation units as
illustrated in \autoref{fig:mass-spring-damper-cosim}.
Simulation unit $S_1$ represents the spring and the damper.
It has one input, $u_1$, which is the velocity with which the spring and damper
are extended; one output, $y_1$, which is the force by which they oppose the
extension; and an internal state, $x_1$, which is the distance by which they
have been extended.
The state equation for $S_1$ is thus
\begin{equation}
    \dot x_1(t) = \tilde u_1(t).
\end{equation}
The tilde denotes extrapolation, which is necessary since each simulation unit
only knows the true value of its input variables at discrete communication
points $t[0], t[1], \dots, t[n]$.
Throughout this article, we shall assume the simplest extrapolation scheme,
which is just to hold the value constant (i.e., zero-order hold):
\begin{equation}
    \tilde u_1(t) = u_1[n],
    \qquad
    t[n] < t \le t[n+1].
\end{equation}
The angle brackets denote discrete time, that is,
$\cdot [n] \equiv \cdot (t[n])$.
Consequently, the output equation for $S_1$ is written
\begin{equation}
    y_1[n] = - k x_1[n] - c \tilde u_1[n].
\end{equation}

Simulation unit $S_2$ represents the mass.
It also has one input, $u_2$, which is the force acting upon it;
one output, $y_2$, which is its velocity; and two internal states,
$x_2$ and $v_2$, which are its displacement and velocity, respectively.
The state-space equations for $S_2$ are thus:
\begin{align}
    \dot x_2(t) &= v_2(t), \\
    \dot v_2(t) &= \frac{1}{m} \tilde u_2(t), \\
    y_2[n] &= v_2[n]. \label{eq:harmonic-oscillator-y2}
\end{align}
The coupling constraints, effected at communication points, are:
\begin{align}
    u_1[n] &= y_2[n] \qquad \text{(velocity)},
        \label{eq:harmonic-oscillator-velocity-constraint} \\
    u_2[n] &= y_1[n] \qquad \text{(force)}.
\end{align}

Given that we use zero-order hold, the state equations for the two simulation
units are easy to solve analytically, and we get:
\begin{align}
    x_1[n+1] &= x_1[n] + u_1[n] \Delta t[n],
        \label{eq:harmonic-oscillator-x1-analytic} \\
    x_2[n+1] &= x_2[n] + v_2[n] \Delta t[n] + \frac{1}{2m} u_2[n] \Delta t[n]^2,
        \label{eq:harmonic-oscillator-x2-analytic} \\
    v_2[n+1] &= v_2[n] + \frac{1}{m} u_2[n] \Delta t[n],
        \label{eq:harmonic-oscillator-v2-analytic}
\end{align}
where $\Delta t[i] \equiv t[i+1] - t[i]$ are the macro time step sizes.

We then simulate this system using two different co-simulation algorithms:
a fixed-step-size algorithm with $\Delta t[i] = \Delta t_\mathrm{fixed} = 0.1$
for all $i$,
and a variable-step-size algorithm.
In the latter case we use the PI control algorithm described by
\textcite{kyllingstad2025error} with an error estimate
based on the \emph{Energy-conservation-based
co-simulation} (ECCO) method \cite{sadjina2017energy}.
The parameters used for the PI controller and error estimator are listed in
\autoref{tab:ecco-parameters} for the sake of completeness and
reproducibility, but the specific choice of algorithm or error estimator is
almost irrelevant here, so we won't dwell further on the details.

\begin{table}
    \centering
    \begin{tabular}{lcc}
        \toprule
        Parameter & Symbol & Value \\
        \midrule
        Initial step size & $\Delta t[0]$ & 0.1 \\
        Proportional gain & $k_P$ & 0.2 \\
        Integral gain & $k_I$ & 0.1 \\
        Minimum step size & $\Delta t_\mathrm{min}$ & $10^{-5}$ \\
        Maximum step size & $\Delta t_\mathrm{max}$ & 0.1 \\
        Maximum relative step size reduction & $\theta_\mathrm{min}$ & 0.2 \\
        Maximum relative step size increase & $\theta_\mathrm{max}$ & 1.2 \\
        \midrule
        Absolute tolerance & $\delta$ & $10^{-6}$ \\
        Relative tolerance & $\sigma$ & $10^{-6}$ \\
        \bottomrule
    \end{tabular}
    \caption{Parameters for the variable-step-size algorithm and error estimator.}
    \label{tab:ecco-parameters}
\end{table}

Since we have the analytic expressions for the subsystems' behaviour between
communication points
(\ref{eq:harmonic-oscillator-x1-analytic}--\ref{eq:harmonic-oscillator-v2-analytic}),
we do not need to perform numerical integration within the simulation units, and
any errors we observe next must entirely be artifacts of the co-simulation.
We also circumvent any issues that might arise from unfortunate interactions
between integration and macro step sizes
(see e.g.\ \textcite{skjong2019numerical}).

\subsection{The puzzle}

\begin{figure}
    \centering
    \includegraphics{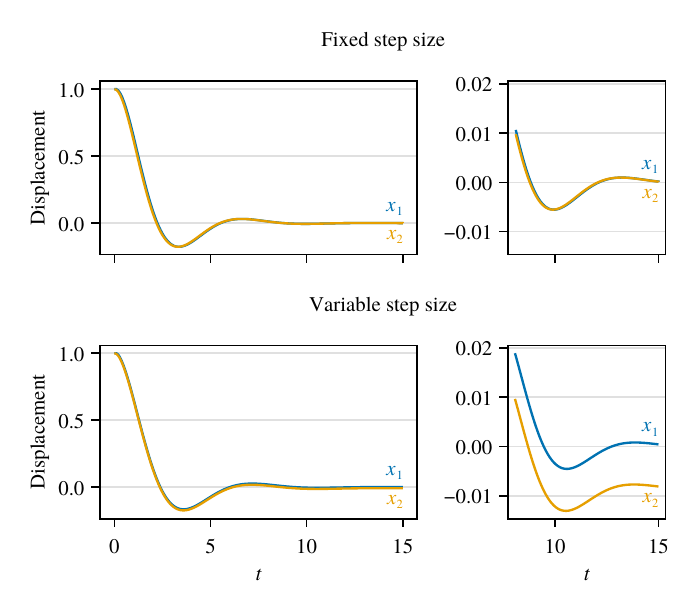}
    \caption{
        Co-simulated solutions of the harmonic oscillator.
        The top panels show the solution obtained with a fixed step size; the
        lower are for the variable-step-size algorithm.
        The leftmost panels show the entire simulation; the rightmost plots show
        a zoomed-in picture of the tail ($9 \le t \le 15$).
    }
    \label{fig:harmonic-oscillator-cosim}
\end{figure}

\autoref{fig:harmonic-oscillator-cosim} shows the results of the two
simulations.
The first thing we note is that the overall results (left-hand panels) are
fairly similar, both to each other and to the accurate solution shown in
\autoref{fig:harmonic-oscillator-accurate}.
However, zooming in on the tail (right-hand panels), we see that there is a
discrepancy between the displacement states $x_1$ and $x_2$ which is especially
prominent in the variable-step-size case.

That there is a discrepancy between the simulation units is no surprise in
itself.
The displacement is obtained through integration of the velocity, and in $S_1$, the
velocity signal from $S_2$ is sampled and inaccurately reconstructed using
zero-order hold.
The surprise is that the difference between $x_1$ and $x_2$ is significantly
larger in the variable-step-size case, despite the fact that we have configured
the controller to use the same or smaller macro step sizes compared to the
fixed-step algorithm ($\Delta t_\mathrm{max} = \Delta t_\mathrm{fixed}$).

\begin{figure}
    \centering
    \includegraphics{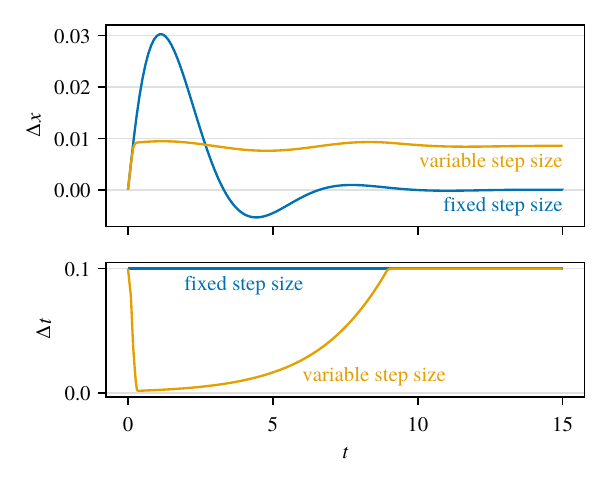}
    \caption{
        The upper panel shows the discrepancy between the displacement states in
        $S_1$ and $S_2$, $\Delta x = x_1 - x_2$, in the fixed-step-size and
        variable-step-size cases.
        The lower panel shows the macro time step size.
        The simulation was run with an initial velocity $\dot x_1 = \dot x_2 = 0$.
    }
    \label{fig:harmonic-oscillator-discrepancy}
\end{figure}

To figure out what is happening here, we look more closely at the discrepancy
between the displacement states in $S_1$ and $S_2$,
\begin{equation}
    \Delta x[n] \equiv x_1[n] - x_2[n],
    \label{eq:harmonic-oscillator-discrepancy}
\end{equation}
alongside a comparison of the step sizes in the two simulations.
This is shown in \autoref{fig:harmonic-oscillator-discrepancy}.
In the fixed-step-size case, we see that the discrepancy initially increases
sharply and varies a lot in the first phase of the simulation.
This is to be expected, since $S_1$ lags behind $S_2$ in terms of its
information about the velocity, which also varies a lot in this phase.
However, the discrepancy vanishes as the system comes to rest.

In the variable-step-size case, we start off with the same initial step size,
and the curves track each other closely in the first couple of time steps.
At a point which coincides with the step size algorithm dropping to a smaller
$\Delta t$, the discrepancy \emph{stops} increasing as sharply and goes
into a phase with much smaller oscillations before it too stabilises.
However, it ends up stabilising at a larger value than in the fixed-step-size
case.
The large discrepancy in the early phase appears to have become ``frozen in'' by
the sudden reduction in step size.

\begin{figure}
    \centering
    \includegraphics{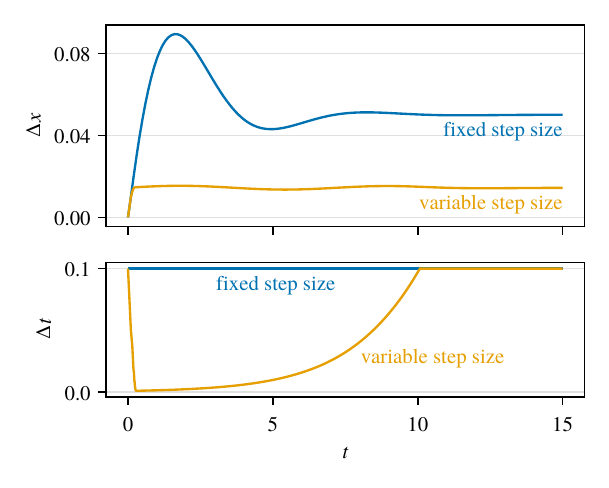}
    \caption{
        The upper panel shows the discrepancy between the displacement states in
        $S_1$ and $S_2$, $\Delta x = x_1 - x_2$, in the fixed-step-size and
        variable-step-size cases.
        The lower panel shows the macro step size.
        Unlike in \autoref{fig:harmonic-oscillator-discrepancy}, the
        simulation was run with an initial velocity $\dot x_1 = \dot x_2 = 1$.
    }
    \label{fig:harmonic-oscillator-discrepancy-v1}
\end{figure}

Apparently, we are observing a kind of numerical error which is somehow linked
to step sizes, though not in the usual way:
Rather than being linked to step size magnitude, and diminishing with smaller
step sizes, it seems to be linked to step size \emph{changes}.
However, it also seems to be heavily dependent on system states, as is shown in
\autoref{fig:harmonic-oscillator-discrepancy-v1}.
Here, we have re-run the simulation with initial velocity
$\dot x_1[0] = \dot x_2[0] = 1$, and now we see that the roles have switched;
the variable-step-size algorithm does better in this case.

\subsection{The explanation}

Note that equations \ref{eq:harmonic-oscillator-x1-analytic}--\ref{eq:harmonic-oscillator-v2-analytic}
for the subsimulators' internal states all take the form of recurrence relations.
Applying them repeatedly down to $n = 0$ yields
\begin{align}
    x_1[n] &= x_1[0] + \sum_{i=0}^{n-1} u_1[i] \Delta t[i], \\
    x_2[n] &= x_2[0] + \sum_{i=0}^{n-1} \left(
        v_2[i] \Delta t[i]
        + \frac{1}{2m} u_2[i] \Delta t[i]^2
        \right), \\
    v_2[n] &= v_2[0] + \frac{1}{m} \sum_{i=0}^{n-1} u_2[i] \Delta t[i].
        \label{eq:harmonic-oscillator-v2-sum}
\end{align}
If we assume that both displacement states are initialised to the same value
($x_1[0] = x_2[0]$), and make use of the output \eqref{eq:harmonic-oscillator-y2}
and constraint \eqref{eq:harmonic-oscillator-velocity-constraint} equations,
which together mean that
\begin{equation}
    u_1[i] = y_2[i] = v_2[i],
\end{equation}
then we find the following compact expression for the state discrepancy
\eqref{eq:harmonic-oscillator-discrepancy}:
\begin{equation}
    \Delta x[n] = -\frac{1}{2m} \sum_{i=0}^{n-1} u_2[i] \Delta t[i]^2.
\end{equation}
This is an example of the kind of extrapolation errors we discussed in the introduction.
The standard ways to deal with it would be:
\begin{itemize}
    \item Transfer higher-order time derivatives of coupling variables.
        For this particular system, if $\dot u_1[n] = \dot y_2[n]$ was
        added to the coupling constraints, the term of order $\Delta t[i]^2$
        would also cancel out entirely.
    \item Lacking that, perform higher-order input extrapolation within $S_1$ to
        obtain \emph{approximations} of $\dot y_2$, $\ddot y_2$, etc.
    \item Reduce the step size, since obviously $\Delta x[n] \to 0$
        quadratically as $\Delta t[i] \to 0$ for all
        $i \in \{0, 1, \ldots, n-1\}$.
\end{itemize}
The last is the simplest, but as we have seen, there is apparently more to that
story.

Consider first the fixed-step-size case where
$\Delta t[i] = \Delta t_\mathrm{fixed} = \mathrm{const}$.
Then, we can write
\begin{equation}
    \Delta x[n]
    = -\frac{1}{2m} \Delta t_\mathrm{fixed} \sum_{i=0}^{n-1} u_2[i] \Delta t_\mathrm{fixed}
    = -\half (v_2[n] - v_2[0]) \Delta t_\mathrm{fixed},
    \label{eq:harmonic-oscillator-fixed-step-discrepancy}
\end{equation}
where we made use of the expression \eqref{eq:harmonic-oscillator-v2-sum} for
$v_2[n]$ in the last step.
This is a dissipative system which eventually comes to rest, so $v_2[n] \to 0$
as $n \to \infty$.
We will therefore end up with a constant error of
\begin{equation}
    \Delta x[\infty] = \half v_2[0] \Delta t_\mathrm{fixed}.
\end{equation}
This is exactly what we see in the upper plots in
figures~\ref{fig:harmonic-oscillator-discrepancy}
and~\ref{fig:harmonic-oscillator-discrepancy-v1}.

Then, we turn to the variable-step-size case, and imagine for a moment that we
only change the step size once, at communication point $K$:
\begin{equation}
    \Delta t[i] = \begin{cases}
        \Delta t_1, & i < K \\
        \Delta t_2, & i \ge K
    \end{cases}
\end{equation}
For $n \ge K$, the displacement discrepancy now becomes
\begin{align}
    \Delta x[n]
    &= -\frac{1}{2m} \Delta t_1 \sum_{i=0}^{K-1} u_2[i] \Delta t_1
        -\frac{1}{2m} \Delta t_2 \sum_{i=K}^{n-1} u_2[i] \Delta t_2 \\
    &= -\half (v_2[K] - v_2[0]) \Delta t_1 - \half (v_2[n] - v_2[K]) \Delta t_2 \\
    &= \half v_2[0] \Delta t_1 + \half v_2[K] (\Delta t_2 - \Delta t_1)
        - \half v_2[n] \Delta t_2.
\end{align}
The last expression is similar to the fixed-step
one~\eqref{eq:harmonic-oscillator-fixed-step-discrepancy}, except there is now
an additional contribution from the step size change which is proportional to
the difference between the two step sizes.
As the simulation runs its course and the system comes to rest, we get a final
discrepancy of
\begin{equation}
    \Delta x[\infty] =
    \half v_2[0] \Delta t_1 + \half v_2[K] (\Delta t_2 - \Delta t_1).
    \label{eq:harmonic-oscillator-discrepancy-infty}
\end{equation}
We can also see this in our results, if we consider only the abrupt step size
change at the beginning of the simulation and ignore the slow rise in step size
thereafter.
In \autoref{fig:harmonic-oscillator-discrepancy} the step size drops by an
order of magnitude at $t[K] \approx 0.2$, so
\begin{equation}
    \Delta t_2 - \Delta t_1 \approx -\Delta t_1 = -0.1.
\end{equation}
At the same point, the velocity (not plotted) is $v_2[K] \approx -0.2$.
Thus, in this case, we get $\Delta x[\infty] \approx 0.1$ which is, to a decent
approximation, what we see in the figure.

For the other initial condition, $v_2[0] = 1$, this single-step-size-change
simplification works less well, possibly due to a larger value of $v_2$ during
the early macro time steps.
In the next section, we will generalise to an arbitrary number of step size
changes.

\section{General case}
\label{sec:general-case}

\begin{figure}
    \centering
    \includegraphics{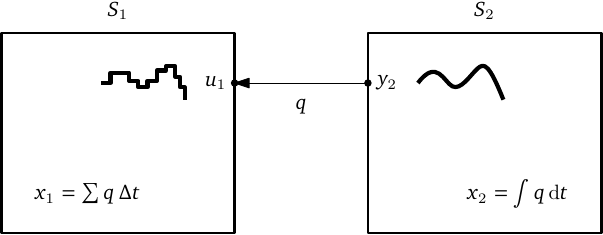}
    \caption{Two systems connected by a flow which accumulates on both sides.}
    \label{fig:general-system-flow}
\end{figure}

Consider now the general case where we have two unspecified systems $S_1$ and
$S_2$ where $S_1$ has an input variable $u_1$ which is connected to $S_2$'s
output variable $y_2$.
(Note that while our first example was based on a closed-loop connection, this
turns out to be beside the point here.)
We shall denote by $q$ the flow that gets transmitted
through this connection, so that, at synchronisation point $n$,
\begin{equation}
    u_1[n] = y_2[n] = q[n] = q(t[n]).
\end{equation}
In both subsystems we have states which represent the integral of $q(t)$,
denoted $x_1$ and $x_2$, respectively.
The setup is illustrated in \autoref{fig:general-system-flow}.

In $S_2$, where $q(t)$ is fully known, the state-space equations are
\begin{align}
    \dot x_2(t) &= q(t), \\
    y_2[n] &= q[n].
\end{align}
In $S_1$, where $q$ is represented by the input variable $u_1$, the
corresponding state equation is
\begin{equation}
    \dot x_1(t) = \tilde u_1(t) = u_1[n] = y_2[n] = q[n],
    \qquad t[n] < t \le t[n+1],
\end{equation}
where we have again extrapolated the input value using zero-order hold.
The value of $x_1$ at communication point $n$ is then
\begin{align}
    x_1[n]
    &= x_1[n-1] + \int_{t[n-1]}^{t[n]} \diff t\; \tilde u_1(t) \nonumber \\
    &= x_1[n-1] + q[n-1] \Delta t[n-1] \nonumber \\
    &= x_1[0] + \sum_{i=0}^{n-1} q[i] \Delta t[i].
\end{align}
The last line is obtained through repeated application of the recurrence
relation in the second line.

We will now derive a similarly structured expression for $x_2$.
We assume that the function $q(t)$ is real analytic at $t[n]$, with the Taylor
series converging to $q(t)$ for all $t \in \left[ t[n], t[n+1]\right]$ and all
communication points $n$.
In other words, we can write
\begin{equation} \label{eq:general_taylor_series}
    q(t) = \sum_{k=0}^\infty \frac{1}{k!} q^{(k)}[n-1] (t - t[n])^k,
    \qquad
    t[n] \le t \le t[n+1],
\end{equation}
where $q^{(k)}[i]$ denotes the $k$\textsuperscript{th} derivative of $q(t)$ at
$t=t[i]$.
Then, it follows that
\begin{align}
    x_2[n]
    &= x_2[n-1] + \int_{t[n-1]}^{t[n]} \diff t\; q(t) \nonumber \\
    &= x_2[n-1] + \sum_{k=0}^\infty \frac{1}{k!} q^{(k)}[n-1]
        \int_{t[n-1]}^{t[n]} \diff t \; (t - t[n-1])^k \nonumber \\
    &= x_2[n-1] + \sum_{k=0}^\infty \frac{1}{(k+1)!} q^{(k)}[n-1] \Delta t[n-1]^{k+1} \nonumber \\
    &= x_2[0] + \sum_{i=0}^{n-1} \sum_{k=0}^\infty \frac{1}{(k+1)!} q^{(k)}[i] \Delta t[i]^{k+1}.
\end{align}
The discrepancy between $x_1$ and $x_2$ (at communication points) can now be
expressed as
\begin{align}
    \Delta x[n]
    &\equiv x_1[n] - x_2[n] \nonumber \\
    &= x_1[0] - x_2[0]
        + \sum_{i=0}^{n-1} \left(
            q[i] \Delta t[i]
            - \sum_{k=0}^\infty \frac{1}{(k+1)!} q^{(k)}[i] \Delta t[i]^{k+1}
        \right) \nonumber \\
    &= \Delta x[0]
        - \sum_{i=0}^{n-1} \sum_{k=1}^\infty \frac{1}{(k+1)!} q^{(k)}[i] \Delta t[i]^{k+1}.
    \label{eq:general_discrepancy_full}
\end{align}

To see more clearly the link to step size changes, we will make two further
assumptions:
firstly, that the states $x_1$ and $x_2$ are given the same initial value, so
that $\Delta x[0] = 0$; and secondly, that all step sizes are so small that
$t[n]$ is sufficiently well within the convergence radius of the Taylor series
of $q(t)$ at $t[n-1]$ that higher-order terms in $\Delta t[i]$ can be neglected.
By the second assumption,
\begin{equation}
    q[n] = q[n-1] + q^{(1)}[n-1] \Delta t[n-1] + \bigO(\Delta t[n-1]^2),
\end{equation}
or
\begin{equation}
    q^{(1)}[n-1] = \frac{q[n] - q[n-1]}{\Delta t[n-1]} + \bigO(\Delta t[n-1]).
\end{equation}
Substituting for this in the $k=1$ term of \autoref{eq:general_discrepancy_full},
the expression for the state discrepancy reduces to
\begin{align}
    \Delta x[n]
    &\simeq - \half \sum_{i=0}^{n-1} \left(q[i+1] - q[i]\right) \Delta t[i] \nonumber \\
    &= \half q[0] \Delta t[0]
        + \half \sum_{i=1}^{n-1} q[i] (\Delta t[i] - \Delta t[i-1])
        - \half q[n] \Delta t[n-1].
    \label{eq:general_discrepancy_simplified}
\end{align}
This shows quite generally that step size changes produce contributions to
the state discrepancy which are, at leading order, proportional to both
the difference in step size from one step to the next and the flow at the
intervening communication point.

\section{Example 2: Connected fluid reservoirs}
\label{sec:connected-reservoirs}

One can argue that the damped harmonic oscillator example we used in
\autoref{sec:harmonic-oscillator} is too pristine and almost seems rigged
to produce the exhibited effect.
The state $x_1$, the displacement as seen from the perspective of the
spring-damper subsystem, is arguably redundant:
It would be natural to treat the displacement of the mass, $x_2$, as the
``ground truth'', and it would be easy to add it as an output from $S_2$ and
take it as an input to $S_1$ instead.

\begin{figure}
    \centering
    \includegraphics{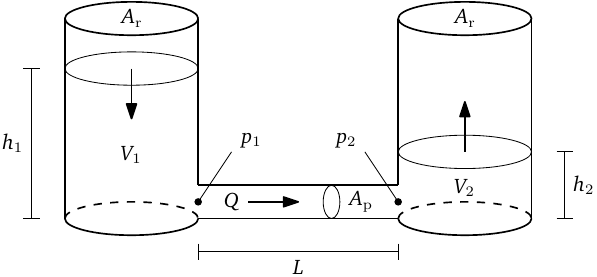}
    \caption{Schematic illustration of the connected fluid reservoirs example.}
    \label{fig:connected-reservoirs}
\end{figure}

Therefore, we close this paper with another example, illustrated in
\autoref{fig:connected-reservoirs},
a model of two fluid reservoirs connected by a pipe.
This is a system where it is quite natural to use a flow variable as the
interface between two simulation units, since the physical manifestation of it
is literally a fluid flow.
It is also an example of a system where, although it would be trivial to expose
the redundant state as an output of $S_2$, it simply makes \emph{less sense} for
the modeller of the individual subsystems to do so:
The integral of the flow is in this case the total amount of fluid which has
passed from one reservoir to the other.
The natural representation of it inside the simulation units, however, is the
total volume of fluid within each reservoir.

We shall employ a highly simplified representation of the system, where the
pressure at the pipe orifices is just the hydrostatic pressure at the bottom of
the reservoirs,
\begin{equation}
    p_i = \rho g h_i = \frac{\rho g V_i}{A_\mathrm{r}},
\end{equation}
and the pressure drop from one end of the pipe to the other is given by the
Hagen--Poiseuille equation for laminar flow of a viscous, Newtonian fluid:
\begin{equation}
    p_1 - p_2 = \frac{8 \pi \mu L Q}{A_\mathrm{p}^2}.
\end{equation}
Physically, this corresponds to a situation where the pipe is very long and
narrow relative to the dimensions of the reservoirs, and the fluid flows
slowly through it.
In these equations, $p_1$ and $p_2$ represent pressure at the bottom of each
reservoir, $V_i$ is the volume of fluid in each of them, and $Q$ is the
volumetric flow rate between them.
$A_\mathrm{r}$ is the area of the reservoirs' horizontal cross sections (assumed
equal), $A_\mathrm{p}$ is the cross-sectional area of the pipe, and $L$ is the
length of the pipe.
$\rho$ and $\mu$ are the density and dynamic viscosity of the fluid,
respectively, while $g$ is the acceleration of gravity.
Since none of the physical details really matter for our purposes, we will, for
the sake of clarity, reduce our notation to only two constants, $C$ and $R$:
\begin{align}
    p_i       &= \frac{V_i}{C}, & C &\equiv \frac{A_\mathrm{r}}{\rho g}, \\
    p_1 - p_2 &= R Q,           & R &\equiv \frac{8 \pi \mu L}{A_\mathrm{p}^2}.
\end{align}
(The symbols for these were chosen by analogy with hydraulic capacitance and
hydraulic resistance, respectively.)

\begin{figure}
    \centering
    \includegraphics{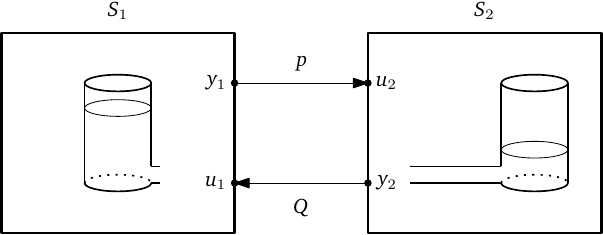}
    \caption{
        Co-simulation setup of the connected fluid reservoirs example.
        Note that $Q$ is positive when fluid flows from reservoir 1
        to reservoir 2 even though we have chosen causality of the connection to
        be in the direction from $S_2$ to $S_1$.
        $p$ is the pressure at the interface between the two subsystems, i.e.,
        at the bottom of reservoir 1.
    }
    \label{fig:connected-reservoirs-cosim}
\end{figure}

Again, we split the system into two simulation units $S_1$ and $S_2$, where the
former represents reservoir 1 and the latter represents reservoir 2 and the pipe.
We connect them via a pressure--flow coupling, where the pressure $p$ is the
pressure at the interface, i.e., at the bottom of reservoir 1.
The setup is shown in \autoref{fig:connected-reservoirs-cosim}.
The state-space equations for $S_1$ are
\begin{align}
    \dot V_1(t) &= -\tilde u_1(t), \label{eq:connected-reservoirs-state-v1} \\
    y_1[n] &= p_1[n] = \frac{V_1[n]}{C},
\end{align}
while for $S_2$ they are
\begin{align}
    \dot V_2(t) &= Q(t) = \frac{\tilde u_2(t)}{R} - \frac{V_2(t)}{C R},
        \label{eq:connected-reservoirs-state-v2} \\
    y_2[n] &= Q[n].
\end{align}
Note here that $y_1$ and $u_2$ represent the interface pressure, $p$, while
$u_1$ and $y_2$ represent the interface flow rate, $Q$.
The coupling constraints are thus
\begin{align}
    u_1[n] &= y_2[n] \qquad \text{(flow rate)}, \\
    u_2[n] &= y_1[n] \qquad \text{(pressure)}.
\end{align}
As in our first example, the state equations
(\ref{eq:connected-reservoirs-state-v1} and
\ref{eq:connected-reservoirs-state-v2}) have straightforward analytical
solutions, so we will not have to worry about errors introduced by their
numerical integration.

We choose parameter values $C = 1$ and $R = 1$ and initial conditions
$V_1(0) = 0.6$ and $V_2(0) = 0.4$.
We use a ``bang-bang'' step size controller which is based on the flow rate:
If the flow is larger than a certain threshold ($Q_\mathrm{t} = 0.5$), then we
use a small step size ($\Delta t_\mathrm{small} = 0.001$), otherwise we use a
large step size ($\Delta t_\mathrm{large} = 0.01$).
That is,
\begin{equation}
    \Delta t[n] = \begin{cases}
        \Delta t_\mathrm{small}, & y_2[n] > Q_\mathrm{t} \\
        \Delta t_\mathrm{small}, & y_2[n] \le Q_\mathrm{t}.
    \end{cases}
\end{equation}
At $t = 1$ we (instantaneously) increase the volume of fluid in $S_1$ by 1
to increase the flow and precipitate a step size change.

\begin{figure}
    \centering
    \includegraphics{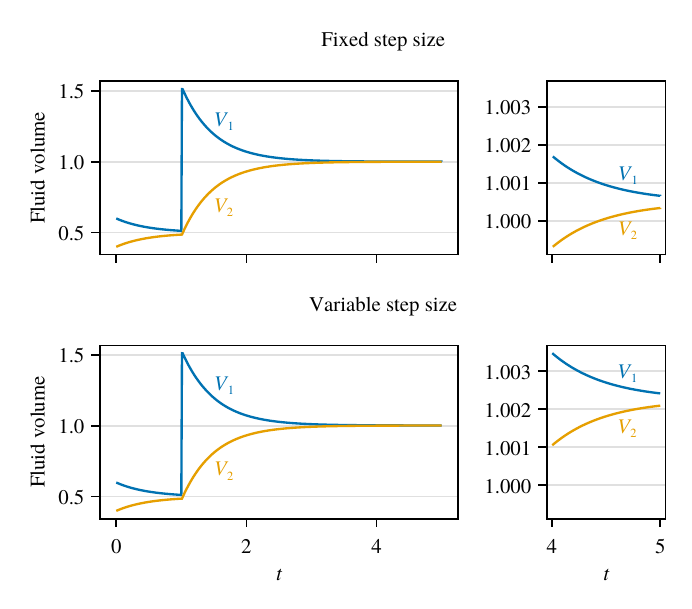}
    \caption{
        Co-simulation results for the connected fluid reservoirs example.
        The top panels show the solution obtained with a fixed step size; the
        lower are for the variable-step-size algorithm.
        The leftmost panels show the entire simulation; the rightmost plots show
        a zoomed-in picture of the tail ($4 \le t \le 5$).
    }
    \label{fig:connected-reservoirs-cosim-result}
\end{figure}

\begin{figure}
    \centering
    \includegraphics{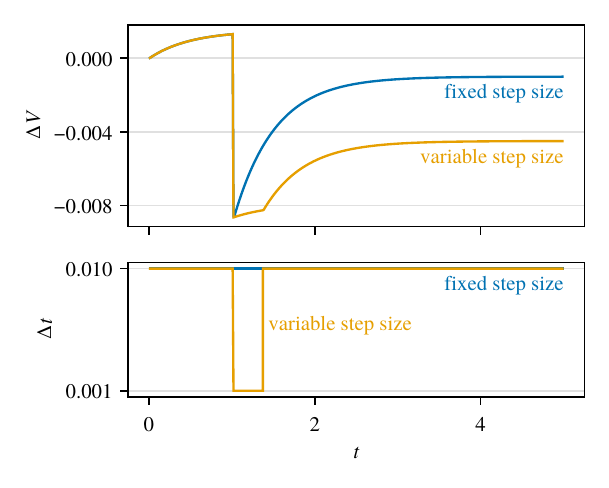}
    \caption{
        The upper panel shows the discrepancy between the displacement states in
        $S_1$ and $S_2$, $\Delta V = V - V_1 - V_2$, in the fixed-step-size and
        variable-step-size cases.
        The lower panel shows the macro time step size.
    }
    \label{fig:connected-reservoirs-discrepancy}
\end{figure}

The results are shown in \autoref{fig:connected-reservoirs-cosim-result}.
Here, we see that the initial imbalance in fluid level between the two
reservoirs decays exponentially, and the fluid volume in both approaches the
equilibrium level at 0.5.
At $t = 1$ we add a unit of fluid to reservoir 1, and after that, the fluid
levels even out at $V_1 \approx V_2 \approx 1$.
As in our first example, we see a difference between the fixed- and
variable-step-size simulations in the zoomed-in plots on the right.
This difference is highlighted in
\autoref{fig:connected-reservoirs-discrepancy}, where we again see that the
discrepancy between the final states is greater in the variable-step-size case.
Here, the discrepancy is calculated as $\Delta V = V - V_1 - V_2$, where $V$ is
the total volume of fluid that has been added to the system (i.e., $V = 1$ for
$t < 1$ and $V = 2$ for $t \ge 1$).

\section{Conclusion}
\label{sec:conclusion}

When two simulation units in a co-simulation are connected via some variable
$q$, and both simulation units have an internal state which represents the time
integral of $q$, there will generally be a discrepancy between those states due
to extrapolation errors.
Such extrapolation errors diminish if the macro time step size is reduced
overall, since the reconstruction of $q$ in one simulation unit will converge
towards the true functional form of $q$ as defined by the other simulation unit.
Conversely, extrapolation errors can be expected to increase with an overall
increase in step size.

What we have shown in this paper, however, is that this picture is more nuanced
when step sizes are changed in a non-uniform manner over time.
A change in step size from $\Delta t$ to $\Delta t'$ at communication point $n$
produces a leading-order contribution to the accumulated discrepancy between the
integral states which is proportional to $q[n] (\Delta t' - \Delta t)$.

This is mainly an issue for variable-step-size simulations where $q$ varies over
time by orders of magnitude.
To understand why this is the case, consider again the general, asymptotic
expressions \eqref{eq:general_discrepancy_simplified} for the state discrepancy:
\begin{align*}
    \Delta x[n]
    &\simeq - \half \sum_{i=0}^{n-1} \left(q[i+1] - q[i]\right) \Delta t[i] \\
    &= \half q[0] \Delta t[0]
        + \half \sum_{i=1}^{n-1} q[i] (\Delta t[i] - \Delta t[i-1])
        - \half q[n] \Delta t[n-1].
\end{align*}
In the first line, it is evident that if $q$ does not vary much, so
$q[i + 1] \approx q[i]$ for all $i$, the accumulated discrepancy will be small.
In the interpretation of the second line, this is because upwards and
downwards step size changes will cancel out over time.
On the other hand, if $q$ varies a lot, and a large step size change happens to
occur when $q$ is large in magnitude, this can incur a discrepancy which does
not necessarily get cancelled out later.
This is exactly what we observed in our two examples.

To guard against such errors, it may help to limit the aggressiveness of the
step size controller in terms of the magnitude of the step size changes it
makes.
For the PI controller we used in \autoref{sec:harmonic-oscillator}, this amounts
to bringing the parameters $\theta_\mathrm{min}$ and $\theta_\mathrm{max}$
closer to 1, or using lower PI gains.\footnote{
    In that particular example, the obvious culprit is the parameter
    $\theta_\mathrm{min} = 0.2$, the value of which was just adopted from the
    original paper \cite{sadjina2017energy}, which allows the step size to be
    reduced by as much as 80\% in a single jump.
}
For simulations which have a highly dynamical initial phase, but which
eventually relax to a steady state, it can help to start the
variable-step-size-controller off at its minimal step size.
This is good general advice when using a feedback step size controller
\cite{kyllingstad2025error}, and especially so in this situation.
Finally, one can imagine developing a step size controller which monitors
flow-type variables and limits step size changes when they are large.
This could be facilitated by a co-simulation interface which supports annotating
variables with value ranges and physical units, such as the Functional Mock-up
Interface (FMI) \cite{modelica2024fmi3}.

We close by pointing out the main limitations of this study, which are that we
have only considered constant input extrapolation (ZOH) and that we used an
explicit, Jacobian co-simulation algorithm.
Both amount to the ``simplest'' choice in each respect, the advantage being that
we were able to carry out a clear and straightforward analysis of the issue.
Since the observed discrepancy is due to extrapolation error, it is to be
expected that higher-order extrapolation and the use of implicit
co-simulation or Gauss--Seidel iteration would reduce the effect, possibly
pushing it to a higher order in $\Delta t$.
This might be an interesting avenue for future research.

\subsection*{Reproducibility}

The code for running the simulations and producing the plots in this paper is
available online at \url{https://doi.org/10.60609/6p8m-0713}.
The code is written in the Julia programming language \cite{bezanson2017julia}
and uses Makie.jl \cite{danisch2021makie} to generate the plots.

\subsection*{Acknowledgements}
This work was supported by the Research Council of Norway under the projects
\emph{SEACo:\ Safer, Easier, and more Accurate Co-simulations} (grant number
326710) and
\emph{OptiStress:\ System optimisation and stress testing in co-simulations}
(grant number 344238).

The author thanks his colleagues Severin Sadjina and Stian Skjong for
interesting discussions and feedback on the work presented herein.

\printbibliography

\end{document}